\documentclass[11pt]{article}

\usepackage{amssymb}
\usepackage{rotating}
\usepackage[greek,english]{babel}
\usepackage{graphicx}
\usepackage{caption}
\usepackage{lscape} 
\usepackage{rotating}
\usepackage{epstopdf}
\usepackage{algorithm}
\usepackage{color}

\input{epsf}

\newcounter{saveeqn}

\newcommand{\bfT}{\mbox{\boldmath $T$}}

\begin{document}

\parindent0mm \parskip0.6cm

%{\Large {\bf Running title: On the relation between clustering and linear modelling} }

%\begin{center}
{\LARGE {\bf Exploring dependence between categorical variables: benefits and limitations of using variable selection within Bayesian clustering in relation to log-linear modelling with interaction terms} }

\begin{large}

MICHAIL PAPATHOMAS

\end{large}

{\it School of Mathematics and Statistics, University of St Andrews, United Kingdom }
%\end{center}

\begin{large}
SYLVIA RICHARDSON
\end{large}

{\it  MRC Biostatistics Unit, Cambridge Institute of Public Health, Cambridge, United Kingdom}

\vspace{0.5cm}

{\bf ABSTRACT: This manuscript is concerned with relating two approaches that can be used to explore complex dependence structures between categorical variables, namely Bayesian partitioning of the covariate space incorporating a variable selection procedure that highlights the covariates that drive the clustering, and log-linear modelling with interaction terms. We derive theoretical results on this relation and discuss if they can be employed to assist log-linear model determination, demonstrating advantages and limitations with simulated and real data sets. The main advantage concerns sparse contingency tables. Inferences from clustering can potentially reduce the number of covariates considered and, subsequently, the number of competing log-linear models, making the exploration of the model space feasible. Variable selection within clustering can inform on marginal independence in general, thus allowing for a more efficient exploration of the log-linear model space. However, we show that the clustering structure is not informative on the existence of interactions in a consistent manner. This work is of interest to those who utilize log-linear models, as well as practitioners such as epidemiologists that use clustering models to reduce the dimensionality in the data and to reveal interesting patterns on how  covariates combine. }

{\it Key words:} Bayesian model selection, sparse contingency tables, graphical models

\section{Introduction}

Detecting high-order interactions is becoming increasingly important for investigators in many fields of research. It is now understood that covariates may combine to affect the probability of an outcome, and that the effect of a particular covariate may only be important in the presence of other covariates. For example, in epidemiology it is of interest to examine the presence of interactions between smoking, environmental pollutants and dietary habits (Bingham and Riboli, 2004). In genetic association studies, it is of interest to detect gene-gene and gene-environment interactions in high dimensional data (Wakefield et al., 2010).

In this manuscript, we examine and discuss the relation between variable selection within Bayesian partitioning on one hand and log-linear modelling with interactions on the other, and the extend to which this relation can be explored in log-linear model search. Log-linear modelling is the most popular approach when searching for interactions, used by statisticians as well as practitioners in substantive applications. In a classical setting, attempting to fit a linear model with a large number of parameters sometimes requires an impractically large vector of observations to produce valid inferences (Burton et al., 2009). Within the Bayesian framework, the use of prior distributions alleviates identifiability or maximum likelihood estimation difficulties; see Dobra and Massam (2010). However, the space of competing models becomes vast, and model search algorithms like the Reversible Jump approach (Green, 1995) require a large number of iteration before they converge and produce reliable posterior model probabilities (Clyde and George, 2004; Dobra, 2009). With regard to contingency tables, the number of cells and possible graphical log-linear models that explain the cell counts increases exponentially with the number of covariates. For example, 
considering 20 covariates with 3 levels implies $3^{20}$ cells and approximately $1.5\times 10^{57}$ possible models.

Due to the difficulties associated with searching for interactions within a linear modelling framework, alternative approaches were adopted focusing on the reduction of the dimensionality in the data. Clustering is often the tool used to reduce dimensionality (see, for example, Zhang et al., 2010), sometimes combined with a variable selection step (Chung and Dunson, 2009).
Whilst log-linear modelling is a standard mathematical construction, there are many different clustering modelling approaches. For the purposes of this manuscript, we choose to focus on Bayesian clustering based on the Dirichlet process. The Dirichlet process produces flexible partitioning, allowing for the evaluation of the uncertainty with regard to the clustering of the subjects. We use a combination of Dirichlet process modelling and variable selection, implementing the modified variable selection step described in Papathomas et al. (2012), so that the covariates that contribute substantially to the clustering are identified. 

We focus on categorical variables and log-linear models, as this is the standard framework for modelling interactions. In fact, for a set of categorical variables, where at least one is binary, there is a correspondence between log-linear and logistic regression modelling, and under certain conditions it is valid to translate inferences from the log-linear framework to the logistic one, regarding the presence of main effects and interactions; see Agresti (2002) and Papathomas (2015, under revision). 

We explore the relation between log-linear modelling and clustering for two reasons. First, practitioners such as epidemiologists often use clustering in order to explore the manner in which covariates combine to affect the risk for disease; see Papathomas et al. (2011a). They frequently question if the clustering structures may inform in some way on the existence of interactions in  associated log-linear models, and our investigation aims to provide some answers. Second, we aim to explore if any relation between log-linear modelling and clustering can be utilized to assist the exploration of large log-linear model spaces and the search for high-order interactions. 
The intuitive idea is that models that combine clustering and variable selection do not select covariates in accordance with the size of their marginal effect. Covariates are selected because they work together and combine with each other to create distinct groups of subjects. Consequently, this type of modelling may be able to inform on covariates that combine to describe the structure in the data, rather than covariates with a strong marginal signal.

In this manuscript, we are not concerned with the large-p problem, where thousands or hundreds of thousands of covariates are considered; see, for example, Hans et al. (2007), Richardson et al. (2010), or Cho and Fryzlewicz (2012) for a comprehensive review. Although our discussion is relevant to data sets of higher dimension, we focus on a relatively modest number of categorical variables, say one hundred or fewer, with fewer than twenty involved in interaction terms. 

We demonstrate that inferences from clustering can potentially reduce the number of factors considered, by determining covariates that are independent of all others. 
Subsequently, the number of competing log-linear models is reduced, making the exploration of the model space feasible. This is crucial when analyzing data that form large sparse contingency tables. We introduce a novel model search approach for a log-linear model space, informed by results from variable selection within clustering. We demonstrate that this model search algorithm can identify parts of the model space that contain models of low probability (thus helping to locate the highest probability model in less iterations, on average, compared to a less informed approach), especially in the presence of covariates that are independent of all other factors. With regard to limitations, first we show that there is no dependable correspondence between the covariate profile of the generated clusters and the log-linear model that best describes the data. More importantly, using simulated and real data, we show that variable selection within Bayesian clustering does not consistently detect marginal independence between covariates when the independent covariates form interaction terms with other factors. 

Studies on the relation between the two different modelling approaches are not commonplace. In Dunson and Xing (2009), a Dirichlet process mixture of product multinomial distributions is defines the prior on a set of categorical variables. Bhattacharya and Dunson (2012) model the joint distribution of categorical variables using simplex factor models. In contrast to our approach, variable selection switches are not considered in the aforementioned manuscripts, and no direct connection is made with log-linear model search. We are aware of three recent manuscripts that utilize clustering. The first is Marbac et al. (2014), where the clustering is applied to the covariates. This is different to the clustering we consider, widely used by practitioners, where the partitioning is applied to the subjects of the study. The second, Johndrow et al. (2014), has some connection to our work. In this preprint, the authors examine situations where the joint distribution implied by a sparse log-linear model has a low-rank tensor factorization. Relevant to our work is also the third, Zhou et al. (2015). This manuscript introduces and utilizes the idea that marginally independent variables reduce the dimensionality of the problem. This approach, central also to our work, was conceived and developed independently in parallel in our manuscript. The modelling in Zhou et al. (2015) with regard to marginal independence has similarities with the one we adopt, and significant differences. Our focus is different from Zhou as we utilize results from clustering to accelerate Bayesian log-linear graphical model selection with the Reversible Jump, a novel approach in log-linear model determination. We come back to these points of comparison in the Discussion Section. 

Section 2, provides a brief description of the clustering and log-linear modelling approaches and contains concepts and notation important to the rest of the manuscript. In Section 3, we present theoretical results on the correspondence between marginal independence on one hand, and variable selection within the Dirichlet process clustering approach on the other, as well as a novel model search approach for log-linear models. Five simulated data sets are analyzed in Section 4, and two real data sets in Section 5. We conclude with a discussion.

\section{Clustering and log-linear models}

\subsection{A Dirichlet process clustering model}

The Dirichlet process (DP) is especially suited to the
problem of clustering observations $x_{1},...,x_{n}$, without
pre-specifying the number of clusters. It is assumed that given parameters $\mu_{i}$, $x_{i}$
is drawn from $F(\mu_{i})$. The mixing distribution over the
parameters $\mu_{i}$ is denoted by $G$. A suitable prior for $G$ is a Dirichlet process with scale
parameter $\alpha$ and mean distribution $G_{0}$. Using $G_0$ and $\alpha$, the DP partitions the $\mu_i$ parameters into a discrete set in a flexible way, allowing the sharing of information between different but similar observations.
Dirichlet process mixture models have been thoroughly investigated in the past (Ferguson, 1973; Lo, 1984; MacEachern and Muller, 1988; Walker et al., 1999; Green and Richardson, 2001). They are used in a wide range of applications, including epidemiology and genetic studies (Huelsenbeck and Andolfato, 2007; Dunson et al., 2008; Sinha et al., 2010; Reich and Bondell, 2011).

We adopt the conjugate Dirichlet process mixture
model used in Molitor et al. (2010) and Papathomas et al. (2011a) for profiling patterns of covariates in epidemiological studies.
For subject $i$, a covariate profile $x_{i}$ is a vector of categorical covariate
values $x_{i}=(x_{i1},...,x_{iP})$, where $P$ is the number of
covariates.
Let $z=\{z_1,...,z_n \}$, where $z_{i}$ is an allocation variable, so that $z_{i}=c$
denotes that subject, $i$, belongs to cluster $c$. Denote with
$\phi^{c}_{p}(x)$ the probability that the $p^{th}$ covariate $x_{.p}$ is equal to
$x$, when the individual belongs to cluster $c$.
Given that
$z_{i}=c$, covariate $x_{.p}$ has a multinomial distribution with
cluster specific parameters
$\phi^{c}_{p}=[\phi^{c}_{p}(1),...,\phi^{c}_{p}(M_{p})]$. Here,
$M_{p}$ denotes the number of categories of $x_{.p}$. We assume that,
a priori, $\phi^{c}_{p}\sim
\mbox{Dirichlet}(\lambda_{1},...,\lambda_{M_{p}})$,
Denote with $\psi = \{ \psi_{c}, \hspace{0.1cm} c\in {\it N}\}$ the probabilities that a subject is
assigned to cluster $c$. We adopt a flexible `stick-breaking' prior on the allocation weights
$\psi_{c}$, with a random parameter $\alpha$ (West, 1992; Ishwaran and James, 2001). For $\phi=\{\phi^{c}_{p}, \hspace{0.1cm} c\in {\it N}, p=1,...,P \}$, the model is written as,

\[
x_i | z,\phi \sim \prod_{p=1}^{P} \phi^{z_i}_{p}(x_{ip}) \mbox{ for } i=1,2,...,n.
\]
\[
\phi^{c}_{p}(x_{ip}) \sim \mbox{Dirichlet}(\lambda_{1},...,\lambda_{M_{p}}) \mbox{ for } c=1,2,...
\]
\[
P(z_i=c|\psi)=\psi_c \mbox{ for } i=1,2,...,n, \mbox{ and } c=1,2,...
\]
\[
\psi_c=V_c \prod_{l<c} (1-V_l) \mbox{ for } c=2,3,... \mbox{ with } \psi_1 = V_1,
\]
\[
V_c \sim \mbox{Beta}(1,\alpha) \mbox{ for } c=1,2,...
\]
This implies the more recognizable mixture for the likelihood of the covariate observations,
\begin{eqnarray}
Pr(x_{i} | \phi,\psi) = \sum_{c=1}^{\infty} \mbox{Pr}(z_{i}=c|\psi) \prod_{p=1}^{P}
 \mbox{Pr}(x_{ip}|z_{i}=c) = \sum_{c=1}^{\infty} \psi_{c} \prod_{p=1}^{P}
\phi^{c}_{p}(x_{ip}). \nonumber
\end{eqnarray}
To identify the covariates that are important for the formation of clusters we consider the
variable selection approach described in Papathomas et al. (2012), which is inspired from Chung and Dunson (2009). In summary, consider cluster specific binary indicators, $\gamma_{p}^{c}$, so that $\gamma_{p}^{c}=1$ when covariate $x_{.p}$ is important for allocating subjects to cluster $c$; otherwise $\gamma_{p}^{c}=0$.
Denote by $\pi_{p}(x_{ip})$ the marginal probability that covariate $x_{.p}$ takes the value $x_{ip}$,  $P(x_{.p}=x_{ip})$. Note that caution should be exercised when interpreting this probability, as it is linked to the sampling frame.  
The probability that covariate $x_{.p}$ is observed as $x_{ip}$, when subject, $i$, belongs to cluster $c$, is written as,
\begin{eqnarray}
P(x_{.p}=x_{ip} \mid z_i=c)=[\phi^{c}_{p}(x_{ip})]^{\gamma^{c}_{p}} \times [\pi_{p}(x_{ip})]^{(1-\gamma^{c}_{p})}.
\end{eqnarray}
Utilizing $\pi_{p}(x_{ip})$ in (1) when the $x_{.p}$ covariate does not contribute to the clustering is intuitively appropriate, as $P(x_{.p}=x | z_i=c)=P(x_{.p}=x)$ implies by Bayes Theorem that $P(z_i=c | x_{.p}=x)=P(z_{i}=c)$. Now, we can write,  
\[
\pi_p(x_{ip})=P(x_{.p}=x_{ip}) = \sum_{c} \psi_c  [\phi^{c}_{p}(x_{ip})]^{\gamma^{c}_{p}} \times [\pi_{p}(x_{ip})]^{(1-\gamma^{c}_{p})}.
\]
We assume that the $\gamma_{p}^{c}$ are independent Bernoulli
variables with $\gamma_{p}^{c} \sim \mbox{Bernoulli} (\rho_{p})$, $0<\rho_{p}<1$. Here, $\rho_p$ describes the probability that covariate $x_{.p}$ is important for the partitioning of the subjects, in relation to the whole process rather than a specific cluster. For $\rho_p$, we consider a sparsity inducing prior with an atom at zero, so that
$\rho_{p} \sim 1_{\{w_{p}=0\}} \delta_{0}(\rho_{p}) + 1_{\{w_{p}=1\}}  \mbox{Beta} (\alpha_{\rho}, \beta_{\rho})$,
where $w_{p} \sim \mbox{Bernoulli}(0.5)$. This prior is appropriate when it is required to clearly discriminate between important and non-important covariates.
The Dirichlet process model described in this Section is fitted using the R package PReMiuM (Liverani et al., 2015).

To create an easily interpretable clustering end-product, whilst the rich MCMC output is utilized and uncertainty is accounted for, we have adopted the model averaging approach described in Papathomas et al. (2012). One aspect of this approach is the derivation of a specific partition that best represents the variable clustering of the subjects during the MCMC run. We refer to this as the `representative partition'.  
To clarify our model and notation we give a simple illustrative example. Consider six categorical covariates, $x_{.1},...,x_{.6}$, taking values 0,1 and 2. Suppose that  subjects are typically allocated into three sub-populations, with probabilities $\psi_{1}=0.3$, $\psi_{2}=0.3$ and $\psi_{3}=0.4$. The multinomial probabilities for the six covariates, given the allocation $z_{i}$ of subject $i$, is given in Table 1a. 
For instance, for $z_{2}=3$ the second subject is allocated to the third group, and the multinomial probabilities for $x_{.1}$ with regard to that subject are, 
$\phi_{1}^{3}=(0.29,0.7,0.01)$. Covariates $x_{.5}$ and $x_{.6}$ clearly do not contribute to the clustering of the subjects, as the multinomial probabilities are the same across clusters. This implies that 
$\gamma_{5}^{c}=\gamma_{6}^{c}=0$ for all $c$. The proportions for the covariate values across the whole sample can be evaluated in accordance with the $\psi_{c}$ and $\phi_{p}^{c}$ parameters. For example, $\pi_{1}(0)=0.3\times 0.01 + 0.3 \times 0.01 + 0.4 \times 0.29=0.122$. For $x_{.5}$ and $x_{.6}$ this evaluation is trivial; for example $\pi_{5}(0)=0.8$. After sampling from this population, a hypothetical summary profile of the three clusters can be derived using the posterior distributions of the model parameters; see Table 1b.  For each covariate $x_{.p}$ and each possible observation $x=0,1,2$, we consider the 95\% credible interval (CI) for the difference between the probability $\phi^{c}_{p}(x)$ of attribute $x$ in group $c$, and the corresponding frequency of $x_{.p}=x$ in the whole sample. Suppose that, with regard to the first group and the first covariate, the two CIs that correspond to $x=0,1$ are both below zero, whilst the CI that corresponds to $x=2$ is above zero. So, for subjects in the first group, it is less likely to observe $0$ or $1$ at the first covariate, compared to the whole sample, and more likely to observe $2$. We denote this information with the `$<$' and `$>$' symbols. We use the `0' symbol when the CI contains zero. In Table 1b, where we also provide hypothetical posterior medians for the selection probabilities $\rho_{p}$, $p=1,...,8$, one can see the hypothetical summary structure in the population. 

\subsection{Log-linear graphical models}

Denote with $\mathcal{P}$ the finite set of the $P$ categorical covariates or factors. The resulting data can be arranged as counts in a $P$-way contingency table. A Poisson log-linear interaction model is a generalized linear model where the data are the cell counts of the contingency table; see Supplemental material, Section S1, for a formal definition of an interaction term in a log-linear model. The number of all possible log-linear models is $2^{(2^{P})}$. It can be very large for non-trivial applications. For example, the number of possible log-linear models for six factors is approximately $184 \times 10^{19}$. Graphical models are a subset of the class of log-linear models. They are represented by a graph where each node (or vertex) is an element of $\mathcal{P}$. Any two nodes may be connected by an edge. Nodes not connected directly by a single edge are independent conditionally on the factors represented by all other nodes (pairwise Markov property). Also, conditionally on nodes to which $x_{.p}$ is directly connected, $x_{.p}$ is independent of all other nodes (local Markov property). Finally, two sets of nodes are independent when they are separated by another set, conditionally on the separating set (global Markov property); see Lauritzen (2011) for more details. The number of possible graphical models is $2^{H}$, where $H=P!/(2(P-2)!)$,
assuming the intercept and all factor main effects are included in the model. For example, the number of possible graphical models for six covariates is 32768.

\section{Results on marginal independence and a novel model search algorithm}

\subsection{Clustering and independence}

%The Theorems below are proven under the assumption that $\pi_{p}(x)=P(x_{.p}=x)$. This assumption holds asymptotically by the Law of Large Numbers, because $\pi_{p}(x)$ is defined as the proportion of times $x_{.p}$ becomes $x$ in the whole sample. This is an additional constraint that is natural and useful for connecting the cluster specific probabilities with the marginal probabilities. 

{\bf Theorem 1:} Consider random variables $x_{.p}$ and $x_{.q}$, $1 \leq p,q\leq P$, $p\neq q$. If $\sum_{c=1}^{C} \gamma_{p}^{c} \times \gamma_{q}^{c} =0$ then $x_{.p}$ and $x_{.q}$ are independent.

{\it Proof:} See Appendix.

{\bf Theorem 2:} Consider a set of random variables $\{x_{.1},\dots,x_{.P} \}$. If, for some $p\in \{1,...,P \}$, $\sum_{c=1}^{C} \gamma_{p}^{c} \times \gamma_{q}^{c} =0$, for all $q\neq p$, then $x_{.p}$ is independent of 
$\{x_{.1},\dots,x_{.P} \} \setminus x_{.p}$.

{\it Proof:} See Appendix.

Note that pairwise independence does not imply independence between sets of random variables. For example, if $x_{.1}$ is independent of $x_{.2}$ and of $x_{.3}$, it is not implied that $x_{.1}$ is independent of $\{x_{.2},x_{.3}  \}$. 
% see Papaiwannou page 65. 
It is also crucial to note that the converse of Theorems 1 and 2 is not necessarily true. The previous Theorems lead to the following Corollary,

{\bf Corollary:} Consider a set of random variables $\{x_{.1},\dots,x_{.P} \}$. If for some $p\in \{1,...,P \}$, $\sum_{c=1}^{C} \gamma_{p}^{c} =0$, then $x_{.p}$ is independent of $\{x_{.1},\dots,x_{.P} \} \setminus x_{.p}$.

Therefore, if the selection probability $\rho_p$ for $x_{.p}$ is zero or close to zero, something that implies that $\sum_{c=1}^{C} \gamma_{p}^{c}$ is also zero or close to zero, we can assume that $x_{.p}$ is not connected with an edge with another covariate. If our interest lies in exploring interactions, to reduce the dimensionality of the problem when fitting log-linear models to sparse contingency tables, $x_{.p}$ could be removed from the analysis.   

\subsection{Construction and interpretation of matrix $\bfT_{\gamma}$}

Considering the results in Section 3.1, we construct $\bfT_{\gamma}$, a matrix that summarizes the variable selection within the clustering output, and translates it into information that is relevant to log-linear modelling. The algorithm for the formation of $\bfT_{\gamma}$ is given below. 

\begin{itemize}

\item For iteration $i_t$ and for each cluster $c$ with more than one subject, form matrix $\bfT^{c,i_t}$, so that element $(p_{1},p_{2})$, $1\leq p_{1} < p_{2}\leq P$ is either zero or one, and equal to  $\gamma_{p_1}^{c}(i_t) \times \gamma_{p_2}^{c}(i_t)$. All other matrix cells are empty.

\item Sum up all matrices $\bfT^{c,i_t}$, weighing by cluster size, to create an information matrix $\bfT_{\gamma}$,
\[
\bfT_{\gamma} = \sum_{i_t} \sum_c n_{c,i_t} \times  \bfT^{c,i_t}.
\]
where $n_{c,i_t}$ is the size of cluster $c$ at iteration $i_t$. Therefore, $\bfT_{\gamma}$ is a straightforward summary of all $\bfT^{c,i_t}$ matrices into one, with small clusters contributing less to this summary.

\item For ease of interpretation reweight the elements of $\bfT_{\gamma}$ so that the maximum element is one,
$\bfT_{\gamma} = (max\{ \bfT_{\gamma} \})^{-1} \times \bfT_{\gamma}$.

\end{itemize}

Matrix $\bfT_{\gamma}$ is constructed in such a manner so that if element $t_{\gamma}(p_{1},p_{2})$, $1\leq p_1 < p_2\leq P$, is close to zero, this implies that an edge between $x_{.p_1}$ and $x_{.p_2}$ is not likely to be present in a highly supported graphical model. 

\subsection{A modified log-linear model search algorithm}

In this subsection, we propose a novel model comparison approach based on the Reversible Jump MCMC algorithm implemented in Papathomas et al. (2011b). We allow for the removal, addition or replacement of one edge in the graph with another. Whilst in the aforementioned manuscript the choice of edge was completely random, we now {\it inform this choice} by the clustering output using $\bfT_{\gamma}$.  

To propose the addition of an edge to the currently accepted model, we consider the elements of $\bfT_{\gamma}$ that correspond to pairs of covariates not currently connected with an edge, transform so that they sum to one, and sample an edge using the derived probabilities. To suggest an edge for removal, we consider the elements of $\bfT_{\gamma}$ that correspond to pairs of covariates already connected with an edge, transform so that they sum to one, and sample an edge using complimentary probabilities. To choose one edge to replace another, we sample both edges as previously. A detailed demonstration of the calculations described in this subsection is presented in the Supplemental material, Sections S2 and S3.

\section{Simulation studies}

The translation we implement between clustering and log-linear model search is novel. We therefore present an extensive range of simulation studies to demonstrate advantages and limitations. The first describes a relatively simple dependence structure. More complex structures are studied in the next two simulations, whilst the last two demonstrate the benefit of our approach with regard to the analysis of sparse contingency tables. 

\subsection{The simulated data sets}

The specifications for the five simulations are shown in Table 2. For simulations 1-3, the majority of the subject observations (80\%) is simulated using Model 1. The rest of the subjects are simulated using Models 2 and 3 in a balanced manner. The models are presented in Figure 1. Simulation 1 is based on two disctinct sets of covariates, where covariates that belong to differnt sets are independent.  Simulations 2 and 3 describe more complex structures compared to simulation 1, since interaction terms share common covariates.  We provide additional information on the design matrices and parameter coefficients of the utilized log-linear models in the Supplemental material, Section S4. We used three models to generate each simulated data set, rather than one, in order to emulate more accurately the variability and complexity within a real data set. 

Two more simulated data sets were created to demonstrate how our approach can be used for the analysis of sparse contingency tables. In simulation 4, only six out of twenty factors are important for explaining the variability associated with the cell counts. In simulation 5, only eight out of 100 factors are important for explaining the variability associated with the cell counts.
Three models were used for the generation of the fourth and fifth simulated data sets, seen in Figure 1, with probabilities \{$032\%, 0.29\%, 0.29\%$\} and  \{$0.8\%, 0.1\%, 0.1\%$\} respectively. 

The size of the model space in simulations 4 and 5 renders conventional model comparison algorithms like the reversible jump MCMC unfeasible.
 The cluster specific variable selection approach should detect that $14$ and $92$ covariates respectively are not important. This will allow for the removal of these covariates from subsequent analyses, forming a drastically smaller model space that can be explored in practice.

\subsection{MCMC specifications, prior distributions and model search strategies}

Information on the size of the chains, as well as run times, is provided in Table 3. 
The log-linear models were fitted and compared within the reversible jump MCMC framework described in Papathomas et al. (2011b). Simulation 4 contains factors with three levels each. Subsequently, models contain, on average, a larger number of parameters compared to the other simulations, resulting in a slower Reversible Jump algorithm. Hence, the relatively small number of iterations. Samples are rather small for accurately estimating  posterior probabilities of less prominent models, in model spaces as large as the ones we consider. However, these chains provide valuable information for the mixing performance of the different reversible jump MCMC algorithms.

The following prior specifications were adopted. For the clustering Dirichlet process model we considered a sparse prior for $\rho_p$ with a point mass at zero (see Section 2.1), to force a clear distinction between the covariates that contribute to the clustering and the ones that do not.
Conjugate Dirichlet priors with $\lambda_{1}=...=\lambda_{M_{p}}=0.5$ were adopted for the $\phi_{p}^{c}$ parameters. Chains were initialized by allocating subjects randomly to ten groups. Initial values for all other model parameters were random. Regarding the log-linear model comparison analyses, unit information priors (Ntzoufras et al., 2003) were adopted for the model parameters. All graphs are equally likely apriori. The majority of the specifications described above are also adopted in the real data analyses presented in Section 5, with differences indicated clearly therein.

Following standard practice when building a reversible jump MCMC chain, in $60\%$ of the iterations, a new set of values for the parameters of the currently accepted model is proposed. A jump to a different graphical model is attempted in $40\%$ of the iterations, where it is equally likely to attempt the addition, removal or replacement of one edge with another.
We compare four model search strategies:
\begin{itemize}
\item[(a)] Uniformly random selection. An unrefined model search strategy where all candidate edges are equally likely to be selected. 
\item[(b)] The cluster specific approach described in Section 3.3. 
\item[(c)] A combination of (a) and (b), where (a) is employed in $30\%$ of the iterations and (b) in $10\%$ of the iterations.
\item[(d)] A balanced combination of (a) and (b) where the two model search approaches are each employed in $20\%$ of the iterations.
\end{itemize}

In all analyses, proposals for the model parameters are derived as in Papathomas, Dellaportas and Vasdekis (2011b), a manuscript where the unrefined model search strategy (a) is adopted. To allow for an intelligible comparison with this standard approach, we refer to the Reversible jump algorithm that employs (a) as the PDV approach using the authors' initials. We do not refer to (a) as PDV when covariates are discarded after implementing the clustering algorithm, because this is not a standard step. Note that parameter proposals could also be constructed following Forster et al. (2012), although the two approaches share many characteristics.

\subsection{Simulation results}

\subsubsection{Variable selection within clustering and marginal independence}

The flexible clustering algorithm discriminated clearly between important and unimportant covariates in all five simulations; see Table 4 for the posterior median selection probabilities $\rho_p$. 
Regarding simulations 4 and 5, the original model space contains $1.5\times 10^{57}$ and $2^{4950}$ graphical models respectively. Implementing the PDV algorithm on such vast model spaces is not feasible, since model comparison would be compromised in terms of convergence and numerical stability. For simulation 4, the variable selection approach described in Section 2.1 correctly reduced the number of covariates to six, after discarding 14 covariates with posterior median selection probabilities less than $0.14$, whilst $E(\rho_p)<0.0045$, $p=7,...,20$. Regarding simulation 5, the number of covariates was correctly reduced to eight, with posterior median selection probabilities for the 92 unimportant covariates equal to zero or less than $0.01$.

\subsubsection{The representative cluster profiles in relation to the presence of interactions}

In most simulations we observe some correspondence between the observed clustering structure and the simulated interactions. However, this correspondence is often blurred, and it is not obvious how to infer and untangle the different interaction terms simply by inspecting the cluster profiles shown in Table 4.  
For simulation 1, three clusters were highlighted in the output summary, as indicated by the patterns of `$>$' and `$<$' (see the end of section 2.1). Clusters 1 and 2 correspond to the simulated `ABCD' and `HIJ' interactions. The posterior median for the selection probability for `C' is only slightly lower than the medians of other important covariates, however `C' does not appear to contribute to the formation of the cluster profiles as strongly as the other important covariates.  Cluster 3 clearly corresponds to the `HIJ' interaction. 
In accordance to the simulation set-up, `E' and `F' have very low selection probabilities. Hence in this simulation, the cluster profile `matches' quite clearly the simulated interactions. 
For simulation 2, five clusters were highlighted in the output. Clusters 2-5 seem to correspond to the `ABCD' and `AFG' interactions,  and the selection probabilities for `H',`I' and `J' are low in accordance with the simulation set-up. 
Two clusters were highlighted in simulation 3. Their profiles seem to correspond to the `ABCD' and `AFG' interactions. The posterior selection probabilities for `H', `I' and `J' are as high as the posterior medians of the other important covariates while that of `E' is small, in accordance with the simulation mechanism.  
With regard to the fourth simulated data set, Table 4 presents results from the flexible clustering analysis in relation to the first six covariates, correctly selected by the clustering algorithm. Six clusters comprise the representative partition, but do not display clear separating patterns suggestive of the existence of specific interactions. This is also the case for simulation 5.  

Overall we see that, although suggestive in some cases, the covariate profiles of representative clusters do not inform conclusively on interaction terms within a log-linear modelling framework. This note of caution is of interest to practitioners that employ clustering approaches, as the relation between covariate profiles and interactions within a linear modelling framework is often a matter of inquiry. 

\subsubsection{The derived $\bfT_{\gamma}$ matrices}

The constructed $\bfT_{\gamma}$ matrices are shown below. We display with bold font the values of elements that correspond to an existing edge in the most probable model; see Section 4.3.4 for posterior model probabilities. 

The $\bfT_{\gamma}$ matrices recover the graph of the most likely model well for Simulations 1 and 2, as expected from our discussion of the representative profiles.  In terms of picking up existing or non-existing edges, it is clear in simulations 1-3 that, overall, smaller weight is given to non-existing edges, compared to existing ones. We also notice a  `spill-over' effect in the $\bfT_{\gamma}$ matrices, with blocks of high valued elements corresponding to important covariates that are not connected in the simulated graph.

In simulations 4 and 5, considering the important covariates, the elements of $\bfT_{\gamma}$ are all large, whether they correspond to an existing edge or not.  This illustrates that the converse of the Theorems in Section 3.1 does not hold. There is no significant difference in the derived $\bfT_{\gamma}$ matrices, when the clustering is performed again on the reduced set of covariates.

Importantly, small elements in the $\bfT_{\gamma}$ matrices {\it always correspond to a non-existing edge}. They never indicate that an existing edge is absent, something that would be detrimental to a model search algorithm.
If the value of an element $t_{\gamma}(p_{1},p_{2})$ is low, say less than 0.1, then it is always the case that the edge between $x_{p_{1}}$ and $x_{p_{2}}$ is absent from the high probability graphical model.  Elements $t_{\gamma}$ that correspond to existing edges are usually much larger, at least one or two orders of magnitude larger compared to elements with a clearly low value.
 These results confirm the correspondence between the two types of structures, the specificity of the pattern of small elements in $\bfT_{\gamma}$, and highlight the potential role of clustering algorithms to assist log-linear model search algorithms. 

\begin{tiny}
 \[
  \bfT_{\gamma}^{sim1}=\left( \begin{array}{lllllllllll}
  & A & B & C & D & E & F & G & H & I & J  \\
  A &  & {\bf .52} & .08 & {\bf .50} & .04 & .02 & .02 & .20 & .27 & .15 \\
  B &  &  & {\bf .45} & 1 & .06 & .04 & .03 & .47 & .64 & .47 \\
  C &  &  &  & {\bf .45} & .02 & .02 & .009 & .12 & .23 & .16  \\
  D &  &	&  & 	& .06 &	.04 &	.03 &	.45	& .65 &	.48 \\
  E &  & 	&	 & 	& 	& .003 & 	.003 &	.03 & .04 & 	.03 \\
  F &  &  &  &  &  &  & .002 &	.02 & 	.03 &	.03 \\
  G &  &  &  &  &  &  &  & .02	& .02 &	.02 \\
  H &  &  &  &  &  &  &  &  & {\bf 0.61} & {\bf .56} \\
  I &  &  &  &  &  &  &  &  &  & {\bf .74}
  \end{array}
  \right)
  \]

  \[
  \bfT_{\gamma}^{sim2}=\left( \begin{array}{lllllllllll}
  & A & B & C & D & E & F & G & H & I & J  \\
  A &  & {\bf .57} & .37 & {\bf .72} & .04 & {\bf .96} & {\bf 1} & .19 & .06 & .05 \\
  B &  &  & {\bf .43} & .36 & .02 & .38 & .27 & .08 & .03 & .02 \\
  C &  &  &  & {\bf .63} & .02 & .24 & .24 & .05 & .03 & .03  \\
  D &  &		& 	& 	& .03 &	.48 &	.54 &	.13	& .04 &	.05 \\
  E & 	& 	& 	& 	& 	& .03 & 	.03 &	.005 & .002 & 	.003 \\
  F &  &  &  &  &  &  & {\bf .83}  & .20 & 	.05 &	.04 \\
  G &  &  &  &  &  &  &  & .18	& .05 &	.04 \\
  H &  &  &  &  &  &  &  &  & .01 & .01 \\
  I &  &  &  &  &  &  &  &  &  & .005
  \end{array}
  \right)
  \]
 \end{tiny}

 \begin{tiny}
 \[
  \bfT_{\gamma}^{sim3}   = \left( \begin{array}{lllllllllll}
  & A & B & C & D & E & F & G & H & I & J  \\
  A &  & {\bf .69} & .15 & {\bf .60} & .06 & {\bf .48} & {\bf .70} & .16 & .21 & .55    \\
  B &  &  & {\bf .46} & .91 & .07 & .52 & .86 & .27 & .38 & .72 \\
  C &  &  &  & {\bf .69} & .02 & .10 & .27 & .14 & .19 & .30  \\
  D &  &		& 	& 	& .07 &	.34 &	.70 &	.27	& .39 &	.66 \\
  E & 	& 	& 	& 	& 	& .03 & 	.06 &	.03 & .03 & 	.06 \\
  F &  &  &  &  &  &  & {\bf .95} &	.18 & 	.44 &	.80 \\
  G &  &  &  &  &  &  &  & .30	& .56 &	1 \\
  H &  &  &  &  &  &  &  &  & {\bf .49} & {\bf .62} \\
  I &  &  &  &  &  &  &  &  &  & {\bf .81}
  \end{array}
  \right), \hspace{0.2cm}
  \bfT_{\gamma}^{sim4}   = \left( \begin{array}{lllllll}
   & A & B & C & D & E & F \\
  A &  & {\bf .95} & {\bf 1} & .77 & .85 & .72    \\
  B &  &  & {\bf .98} & .77 & {\bf .83} & .72  \\
  C &  &  &  & .78 & .87 & .73   \\
  D &  &		& 	& 	& {\bf .72} &	.66  \\
  E & 	& 	& 	& 	& 	& {\bf .69}
  \end{array}
  \right)
  \]
 \end{tiny}

 \begin{tiny} 
\[
  \bfT_{\gamma}^{sim5}=\left( \begin{array}{lllllllll}
    & A & B & C & D & E & F & G & H \\
 A &  & {\bf .99} & {\bf 1} & {\bf 1} & 1 & 1 & 1 & 1 \\
 B &  &  & .97 & .99 & .99 & .99 & .99 & .99  \\
 C &  &  &  & .99 & .99 & .99 & .99 & .99   \\
 D &  &		& 	& 	& .99 &	1 &	1 &	1	 \\
 E & 	& 	& 	& 	& 	& {\bf 1} & {\bf 1} &	{\bf 1}  \\
 F &  &  &  &  &  &  & {\bf 1} & {\bf 1}  \\
 G &  &  &  &  &  &  &  & {\bf 1}	
  \end{array}
  \right)
  \]
 \end{tiny}

\subsubsection{Log-linear model selection with the aid of the clustering output}

Due to the relatively small number of subjects in relation to the number of cells in the contingency tables, and the variability inherent in such simulations, posterior model probabilities are not 80\%, 10\% and 10\% for Models 1,2, and 3 shown in Figure 1. In Figure 2, the top 3 models aposteriori as well as model probabilites are presented for each simulation. For simulations 1 to 4, the most likely model aposteriori is the same as the main model used to create the data (Model 1 in Figure 1), whilst this is not the case for simulation 5. Model probabilities were derived using the Reversible jump algorithm and search strategy (d); see results presented in Table 5. 

Simulations 1 to 3 generate contingency tables that are not sparse. The Reversible Jump algorithm can explore the whole set of possible graphical models without removing any covariates from the analysis. 
In contrast, with regard to simulation 4, the removal of 12 marginally independent covariates reduced the size of the contingency table from $3.4\times 10^9$ to $729$ cells, and the number of log-linear graphical models from $1.5\times 10^{57}$ to a more manageable 32768. We performed model comparison on the reduced data set with six covariates, using variation (a) where all proposed moves are random, in effect a variation that corresponds to using PDV after reducing the model space with the cluster specific approach. We also consider the three model search variations that utilize $\bfT_{\gamma}$, (b), (c) and (d).  

Removing 92 marginally independent covariates from the simulation 5 analysis reduced the size of the contingency table from $1.27\times 10^{30}$ to $256$ cells, and the number of log-linear graphical models from $2^{4950}$ to $2^{8}$; a huge gain. Although simulation 5 mainly illustrates the utility of clustering output in reducing the number of covariates for sparse contingency tables, it also illustrates the fact that the converse of Theorem 1 does not hold. For the covariates kept in the analysis, all weights in the $\bfT_{\gamma}$ matrix are effectively equal to one, even for non-existing edges. Consequently, after removing the unimportant covariates, it is not possible to improve on the standard search algorithm by considering the cluster specific output. In fact, model comparison on the reduced data set was performed using only one search strategy, as all four strategies are equivalent. In general, if there is little variability in the elements of the $\bfT_{\gamma}$ matrix, we do not expect that this matrix will be informative to the model search.

In Table 5, we present results on the performance of the different reversible jump chains and search strategies. The cluster specific approach (b) outperforms the other search strategies, in terms of iterations to best model. This effect is more prominent in simulations 1 and 2.
Search strategy (b) offers a noticeably lower acceptance rate in simulation 1, where we observe a trade-off between acceptance rate and number of iterations to the best model.
Intuitively, by having more targeted moves, the overall chance of jumping decreases, but the chain moves more quickly to the higher posterior probability region.

Overall, results in simulations 1 to 3 show the benefit of search strategy (b), where information from variable selection within clustering is included in log-linear model search. With regard to simulation 4, there is little improvement when the $\bfT_{\gamma}^{sim4}$ matrix is employed; see Table 5. This was expected, as there is little variability in the elements of $\bfT_{\gamma}^{sim4}$. In the Supplemental material, Section S5, we examine the rate of accumulated mass of posterior model probability for the first 3 simulations and the different search strategies employed. The reported results also support the argument for incorporating information from variable selection within clustering. 

Although our experimental results support search strategy (b), strategy (d), where (a) is combined in a balanced manner with (b), also performs well, 
offering a good balance between acceptance rate and iterations to best model. 
Although we did not observe this in any of our analyses, it is prudent to include random search steps that do not depend on the derived $\bfT_{\gamma}$ matrix as a safeguard, in case variable selection within clustering does not detect an existing edge in a high probability graphical model. In this hypothetical scenario, the search moves that do not depend on $\bfT_{\gamma}$ will allow for the detection of the covariate space that is not supported by the clustering. Note that edges not reflected in $\bfT_{\gamma}$ are likely to exist in lower probability models. 

\section{Real data illustrations}

MCMC specifications for the two real data illustrations, as well as run times, are given in Table 3. Prior distributions were the same as the ones adopted in the analysis of the simulated data, described in Section 4.2. 

\subsection{Risk factors for coronary heart disease}

Edwards and Havr\'anek (1985) presented a $2^{6}$ contingency table
in which $1841$ men were cross-classified by six risk factors for
coronary heart disease (CHD). We assume
that main effects are always present and compare the $32768$ possible graphical log-linear
models. Due to the large number of times this data set has been analyzed in the past [see, for example, Dellaportas and Forster (1999)] the top two graphical models (`ADE+AC+BC+BE+F' and `AE+DE+AC+BC+BE+F', following the notation in Agresti, 2002) and associated posterior model probabilities (0.28 and 0.23 respectively for unit information priors) are known. All other graphical models have posterior probabilities lower than 0.1.

In Table 4, we present the covariate profiles of the representative clusters created with the Bayesian partitioning analysis. The subjects are divided in two clusters, and it is not straightforward to disentangle the log-linear model interactions that are present from the cluster profiles. 

The two-way interactions `AC', `AE', `BC' and `BE' are clearly captured by $\bfT_\gamma$; see below. As in Section 4.3.3, we display with bold font the values of elements that correspond to an existing edge in the most probable model. This demonstrates the applicability of our approach. Elements $t_{\gamma}(1,4)=0.14$ and $t_{\gamma}(4,5)=0.12$ that correspond to the three-way interaction `ADE' are smaller. We believe this is due to the signal in the data not being strong. The two likely models have combined posterior probability equal to 0.51, whilst only one of them contains the three-way interaction `ADE'. No other model is associated with probability greater than 0.1.  Nevertheless, the two elements $t_{\gamma}(1,4)$ and $t_{\gamma}(4,5)$ are still one order of magnitude larger compared to the five elements that correspond to `F'. Factor `F' does not interact with any other covariate, and this matches the low posterior selection probability $E(\rho_{6})=0.10$, implying it is not likely to propose the addition of an edge in the graphical model from covariate `F' to another covariate. Of the eleven edges that are not present in the high probability model, five correspond to very small elements $t_\gamma$. Using $\bfT_{\gamma}$ to inform the model search algorithm, results in the identification of a large part of the model space that is associated with low probability.

\begin{tiny}
 \[
  \bfT_{\gamma}^{\mbox{Real data (CHD)}}   = \left( \begin{array}{lllllll}
    & A & B & C & D & E & F \\
 A & & .81 & {\bf .81} & {\bf .14} & {\bf .56} & .04    \\
 B & &  & {\bf 1} & .16 & {\bf .75} & .05  \\
 C & &  &  & .16 & .75 & .05   \\
 D & &		& 	& 	& {\bf .12} &	.01  \\
 E &	& 	& 	& 	& 	& .05
  \end{array}
  \right)
  \]
 \end{tiny}

In Table 6, we present model selection results. It is clear that adopting search strategy (b) to incorporate information from the clustering analysis reduces the average number of iterations to the best model.
Model search strategy (d), where (a) and (b) are combined also performs well, as was the case in the simulations.

\subsection{Genetic and other risk factors}

We consider thirty single nucleotide polymorphisms (SNPs) in chromosomes 6 and 15. These are data from 4260 subjects that participated in a genome-wide association study of lung cancer presented in Hung et al. (2008). The thirty most significant SNPs in terms of marginal p-value are analyzed. Some of these genetic markers were identified as associated with the phenotype in Papathomas et al. (2012). We consider two levels for each marker (0-wild type; 1- homozygous or heterozygous variant).

Twelve SNPs were indicated as important by variable selection within clustering; two from chromosome 15 and ten from chromosome 6. Nine of the selected chromosome 6 SNPs are highly correlated. The two selected chromosome 15 SNPs are also highly correlated. Therefore, we decided to include three SNPs in the log-linear graphical model as representatives of the selected SNPs; rs8034191 from chromosome 15 and \{rs4324798,rs1950081\} from chromosome 6. We also include age, gender and smoking status in the log-linear graphical model, to search for gene-environment interactions as well as gene-gene interactions. We consider two levels for smoking (0-non or ex smoker; 1- smoker) and age (below and above median). The variables will be referred to as A to F, with \{A,B,C\} denoting the genetic factors.

Reducing the number of SNPs from 30 to 12, and then to 3, allows for the use of reversible jump MCMC to compare competing graphical models. The $2^{33}$ contingency table would be too sparse with the vast majority of cells equal to zero.

The highest posterior probability model is `A+B+C+DEF', which does not support the presence of gene-gene or gene-environment interactions. On the other hand, a three-way interaction `DEF' is suggested, which implies different patterns of smoking behaviour by age and gender. The presence of such an interaction is in line with epidemiological understanding, and shows that our algorithm performs well. In Table 4, we present the profiles of the representative clusters created with the Bayesian partitioning analysis. The subjects are typically divided in two clusters which correspond to the `DEF' interaction. 

The derived $\bfT_{\gamma}$ matrix is shown below, after the first stage clustering analysis is performed afresh for the six covariates. We did not cluster the subjects using all 12+3 covariates because the 12 highly correlated important SNPs would `swamp' the 3 environmental factors. The $\bfT_{\gamma}$ matrix correctly indicates the presence of the three-way interaction `DEF'. It also correctly indicates that the first three covariates do not form any interaction terms. In this case, we do see a close correspondence between the clustering pattern and interactions in the associated log-linear model.

\begin{tiny}
 \[
  \bfT_{\gamma}^{\mbox{Real data (GE) (2nd run)}}   = \left( \begin{array}{lllllll}
    & A & B & C & D & E & F \\
 A & & 0.002 & .01 & .06 & 0.06 & .06    \\
 B & &  & .001 & .02 & 0.02 & .02  \\
 C & &  &  & .09 & .07 & .08   \\
 D & &		& 	& 	& {\bf 1} &	{\bf .98}  \\
 E &	& 	& 	& 	& 	& {\bf .88}
  \end{array}
  \right)
  \]
 \end{tiny}

Similarly to the previous real data analysis, using the $\bfT_{\gamma}$ matrix to inform the model search algorithm results in the identification of part of the model space that is associated with low probability and improvement in model search (Table 6). 

We also investigated an alternative approach for assessing the evidence for the presence of an edge, where the pairwise association between two factors is evaluated by the estimation of odds-ratios. See the Supplemental material, Section S6, for more details on these calculations, as well as an illustration on the two real data sets analysed in this manuscript. Results demonstrate that our approach, based on a clustering procedure that considers all variables simultaneously, gives different information on the presence of interactions (two-way and higher) than an approach which is based purely on pairwise associations. For example, for the genetic data analysed in this subsection, the pairwise approach fails to capture an association between D and F, despite the three-way interaction `DEF' present in the prominent highest posterior probability model; see Table S2 in the Supplemental material. We further discuss this in the next Section. 

\section{Discussion}

The advantage in utilizing variable selection within partitioning to inform log-linear model selection is mostly pertinent to marginal independence. For sparse contingency tables, this information can lead to the substantial reduction of the number of covariates considered, making the exploration of the model space feasible. For example, in the second real data illustration, it would be impossible to explore the model space for a $2^{33}$ contingency table by conventional methods such as the  Reversible jump MCMC, without the considerable reduction in the number of SNPs through the first clustering stage. Theoretical results presented in Section 3.1 show that covariates $x_{.p}$ with posterior median  selection probability $\rho_p$ equal to zero (or very close to zero in practice) do not form interaction terms. This appears to be true even when a sparse prior distribution is adopted for the selection parameters $\rho_p$, as was the case for all simulation studies and real data analyses in this manuscript.

With regard to detecting conditional independence, utilizing the output from a clustering model, where all variables are considered simultaneously, offers different results compared to methods based on pairwise associations for the detection of edges. This was illustrated empirically on the two real data examples; see results presented in the Supplemental material. Intuitively, considering all variables simultaneously, rather than in a pairwise fashion, should increase the likelihood of detecting dependence structures that are more complex than pairwise dependencies such as two-way interactions. Nonetheless, it is possible that incorporating in some manner information coming from odds-ratios could be beneficial, given that multiple testing concerns are addressed. Note that our approach utilizes a variable selection approach where all factors are included simultaneously in the model, with a prior assigned to the probability of inclusion. This makes it less susceptible to multiple testing concerns, and particularly suitable for reducing the search space in cases where a large number of factors is investigated; see Scott and Berger (2010). 

Adopting search strategy (b) and informing the model search algorithm with $\bfT_{\gamma}$ often improves the efficiency of the search. Although marginal independence was not always detected, because the converse of the Theorems in 3.1 does not hold, in the majority of the analyses $\bfT_{\gamma}$ identified parts of the model space that contained models of low probability, leading to more efficient model search steps. 
Importantly, using $\bfT_{\gamma}$ to assist the model search never resulted in a worse algorithm, compared to the standard model search approach in Papathomas et al. (2011b). In terms of number of iterations to the best model, the model search algorithm that is informed by clustering performed better or at least as efficiently as the standard algorithm.  The additional computational cost for the clustering is minimal when the R package PReMiuM is used (Liverani et al. 2015), which is primarily written in C++ and R; see the run times reported in Table 3. 
The approach where the naive model search (a) is combined in a balanced manner with (b), where the $\bfT_{\gamma}$ matrix is employed, also performs well, offering a good balance between acceptance rate and number of iterations to the best model. Combining a `naive' with a more `targeted' search approach ensures a comprehensive and efficient exploration of the model space, in the same spirit as the simultaneous sampling from `hot' and `cold' chains in simulated tempering (Geyer and Thompson 1995).

In Johndrow et al. (2014), the authors consider standard and novel latent class structures. The DP is a special case, and its rank is defined as the minimum number of clusters required to describe the joint probability tensor for the categorical covariates. The authors relate log-linear modelling with latent class modelling, investigating if a trivial relationship exists between the two modelling approaches, as we do in this manuscript, albeit from a different standpoint. Bounds are derived for the rank of the latent class model, in relation to the number and structure of the interactions that are present in a weakly hierarchical log-linear model. In one of the results, a massive reduction in the upper bound of the latent class model's rank is shown, under a sparse log-linear model; a model is defined as sparse when the number of non-zero model parameters is much smaller compared to the number of parameters in the saturated model. The authors also demonstrate that the rank of the latent structure depends only on variables that are not marginally independent. A straightforward application of one of the results in Johndrow et al. (2014), gives that an upper bound of the rank of the latent class model corresponding to the prominent model of simulation 1 is $2^7$, rather than the default $2^9$. The upper bound corresponding to the prominent model of simulation 5 is $2^8$, rather than the default $2^{99}$. 

Zhou et al. (2015) also utilizes the idea that marginally independent variables reduce the dimensionality of the model required to describe the joint probability distribution between the covariates. A PARAFAC factorization is adopted, which can be viewed as a more general representation of the Dirichlet process. Dimensionality reduction is achieved with the introduction of the sparse PARAFAC (sp-PARAFAC) formulation,  where marginal independence is modelled with fixed baseline vectors, quantities that correspond to the $\pi_{p}(x)$ quantities we introduced in this manuscript. These are the main similarities between the two approaches, although there are significant differences too. In Zhou et al. (2015) the focus of the theoretical results are in providing expressions for parameters of the log-linear models  that correspond to the adopted latent class model, assessing the level of induced shrinkage, and assessing the convergence of the probability tensor induced by the sp-PARAFAC formulation to the true probability tensor. In contrast, we focus our theoretical investigation on the variable selection switches and what they imply with regard to marginal independence. The prior formulation for detecting marginally independent covariates and reducing dimensionality is also different in the two approaches. Finally, the objectives in the two manuscripts are different, as we focus on accelerating log-linear model selection with the Reversible Jump approach by utilizing output from the clustering process.

A limitation of the approach introduced in this manuscript, as well other approaches we discussed, is the inability to detect conditional independence through the clustering output in a consistent and wieldy manner. One recent attempt at tackling this problem is Kunihama and Dunson (2014), where the concept of mutual information is introduced. Results similar to the ones in Section 3.1, concerning conditional independence, would be useful as conditional independence between variables is key when building the joint distribution of $\{x_{.1},\dots,x_{.P} \}$ using graphical models.  Investigating a possible direct link between variable selection within clustering and conditional independence is the subject of ongoing research.

\vspace{0.2cm}
\noindent {\bf Acknowledgements}
\vspace{0.1cm}

\noindent
We would like to thank the Editor, Associate Editor and three reviewers for comments that improved this manuscript. We would also like to thank Professor Paolo Vineis and Dr Paul Brennan for providing the data used in Section 5.2. This work was supported by MRC grant G1002319.

Supporting information. Additional information for this article is available online.  %\begin{verbatim} SuppInfor_Papathomas_Richardson.pdf \end{verbatim}

\parindent 0mm \parskip 0cm
\vspace{0.3cm}
\noindent {\bf References}
\vspace{0.0cm}

\begin{list}{}
{\setlength{\itemsep}{0cm}
\setlength{\parsep}{0cm}
\setlength{\leftmargin}{0.5cm}
\setlength{\labelwidth}{0.5cm}
\setlength{\itemindent}{-0.5cm}
}

\item Agresti, A., 2002. Categorical data analysis. 2nd edition. John wiley \& Sons. New Jersey. 

\item Bingham, S., Riboli, E., 2004. Diet and Cancer - the European prospective Investigation into cancer and nutrition. Nature Reviews. Cancer. 4, 206-215.

\item Bhattacharya, A., Dunson, D.B., 2012. Simplex factor models for multivariate unordered categorical data. J. Am. Stat. Assoc. 107, 362-77.

\item Burton, P.R., Hansell, A.L., Fortier, I., Manolio, T.A., Khoury, M.J., Little, J., Elliot, P., 2009. Size matters: just how big is BIG? Quantifying realistic sample size requirements for human genome epidemiology. Int. J. Epidemiol. 38, 263-273.

\item Cho, H., Fryzlewicz, P., 2012. High dimensional variable selection via tilting. J. Roy. Stat. Soc. B.  74, 593-622.

\item Chung, Y., Dunson, D.B., 2009. Nonparametric Bayes conditional distribution modelling with variable selection.  J. Am. Stat. Assoc. 104, 1646-60.

\item Clyde, M., George, E.I., 2004. Model uncertainty. Stat. Sci. 19, 81-94.

\item Dellaportas, P., Forster, J.J., 1999. Markov chain Monte Carlo model determination for hierarchical and graphical log-linear models. Biometrika, 86, 615-633.

\item Dobra, A., 2009. Variable selection and dependency networks for genomewide data. Biostatistics, 10, 621-639.

\item Dobra, A., Massam, H., 2010. The mode oriented stochastic search (MOSS) algorithm for log-linear models with conjugate priors. Statistical Methodology, 7, 240-253.

\item Dunson, D.B., Herring, A.H., Engel, S.M., 2008. Bayesian selection and clustering of polymorphisms in functionally-related genes. J. Am. Stat. Assoc. 103, 534-546.

\item Dunson, D.B., Xing C., 2009. Nonparametric Bayes modelling of multivariate categorical data. J. Am. Stat. Assoc. 104, 1042-1051.

\item Edwards, D., Havr\'anek, T., 1985. A fast procedure for model search in multi-dimensional contingency tables. Biometrika, 72, 339-351.

\item Ferguson, T.S., 1973. A Bayesian analysis of nonparametric problems. Ann. Stat. 1, 209-230.

\item Forster, J., Gill, R., Overstall, A., 2012. Reversible jump methods for generalised linear models and generalised linear mixed models. Stat. Comput. 22, 107-120.

\item Geyer, C.J., Thompson, E.A., 1995. Annealing Markov chain Monte Carlo with
applications to ancestral inference. J. Am. Stat. Assoc. 90,  909-920.

\item Green, P.J., 1995. Reversible jump MCMC computation and Bayesian model determination. Biometrika, 82, 711-732.

\item Green, P.J., Richardson, S., 2001. Modelling heterogeneity with and without the Dirichlet process. Scand. J. Stat. 28, 355-75.

\item Hans, C., Dobra, A., West, M., 2007. Shotgun Stochastic Search for `Large p' Regression. J. Am. Stat. Assoc. 102, 507-516.

\item Huelsenbeck, J.P., Andolfatto, P., 2007. Inference of population structure under a Dirichlet process model. Genetics, 175, 1787-1802.

\item Hung, R.J., McKay, J.D., Gaborieau, V., Boffetta, P., Hashibe, M., Zaridze, D. et al. 2008. A susceptibility locus for lung cancer maps to nicotinic acetylcholine receptor subunit genes on 15q25. Nature, 452, 633-37.

\item Ishwaran, H.,James, L., 2001. Gibbs sampling methods for stick-breaking priors. J. Am. Stat. Assoc. 96, 161-73.

\item Johndrow, J.E., Bhattacharya, A., Dunson, D.B., 2014. Tensor decompositions and sparse log-linear models. arXiv:1404.0396v1. 

\item Kunihama, T., Dunson, D., 2014. Nonparametric Bayes inference on conditional
independence, arXiv: 1404.1429v1

\item Lauritzen, S.L., 2011. Elements of graphical models. Lectures from the XXXVIth International Probability Summer School in St-Flour, France. http://www.stats.ox.ac.uk/~steffen

\item Liverani, S., Hastie, D. I., Papathomas, M., Richardson, S., 2015. PReMiuM: An R package for Profile Regression Mixture Models using Dirichlet Processes. Journal of Statistical Software, 64, Issue 7.  

\item Lo, A.Y., 1984. On a class of Bayesian nonparametric estimates. I. Density estimates. Ann. Stat. 12, 351-357.

\item MacEachern, S.N., M\"uller, P., 1998. Estimating mixture of Dirichlet process models. J. Comput. Graph. Stat. 7, 223-38.

\item Marbac, M., Biernacki, C., Vandewalle, V., 2014. Model-based clustering for conditionally correlated categorical data. arXiv:1401.5684v2

\item Molitor, J., Papathomas, M., Jerrett, M., Richardson, S., 2010. Bayesian profile regression with an application to the National Survey of Children's Health. Biostatistics, 11, 484-98.

\item Ntzoufras, I., Dellaportas, P., Forster, J.J., 2003. Bayesian variable and link determination for generalized linear models. J. Stat. Plan. Infer. 111, 165-180.

\item Papathomas, M., 2015. On the correspondence between Bayesian log-linear and logistic regression models with $g$-priors. Under revision 

\item Papathomas, M., Molitor, J., Riboli, E., Richardson, S., Vineis, P., 2011a. Examining the Joint Effect of Multiple Risk Factors Using Exposure Risk Profiles: Lung Cancer in Non-Smokers. Environ. Health Persp. 119, 84-91.

\item Papathomas, M., Dellaportas, P., Vasdekis, V.G.S., 2011b. A novel reversible jump algorithm for generalized linear models. Biometrika, 98, 231-236.

\item Papathomas, M., Molitor, J., Hoggart, C., Hastie, D., Richardson, S., 2012. Exploring data from genetic association studies using Bayesian variable selection and the Dirichlet process: application to searching for gene-gene patterns. Genet. Epidemiol. 36, 663-674.

\item Reich, B.J., Bondell, H.D., 2011. A spatial Dirichlet process mixture model for clustering population genetics data. Biometrics, 67, 381-90.

\item Richardson, S., Bottolo, L., Rosenthal, J.S., 2010. Bayesian models for sparse regression analysis of high dimensional data. Bayesian Statistics, 9, 539-569.

\item Scott, J.G., Berger, J.O., 2010. Bayes and Empirical Bayes multiplicity adjustment in the variable selection problem. The Annals of Statistics, 38, 2587-2619.

\item Sinha, S., Mallick, B.K., Kipnis, V., Carroll, R.J., 2010. Semiparametric Bayesian analysis of nutritional epidemiology data in the presence of measurement error. Biometrics, 66, 444-54.

\item  Wakefield, J., De Vocht, F., Hung, R.J., 2010. Bayesian mixture modelling of gene-environment and gene-gene interactions. Genet. Epidemiol. 34, 16-25.

\item Walker, S., Damien, P., Laud, P., Smith, A., 1999. Bayesian nonparametric inference for random distributions and related functions (with discussion). J. Roy. Stat. Soc. B, 61, 485-527.

\item West, M., 1992. Hyperparameter estimation in Dirichlet process mixture models. Institute of Statistics and Decision Sciences.

\item  Zhang, W., Zhu, J., Schadt, E.E., Liu, J.S., 2010. A Bayesian partition method for detecting Pleiotropic and Epistatic eQTL modules. PLoS Comput. Biol. 6, 1-10.

\item Zhou, J., Bhattacharya, A., Herring, A.H., Dunson, D.B., 2015. Bayesian factorizations of big sparse tensors. J. Am. Stat. Assoc. Accepted manuscript. 

\end{list}

\vspace{0.2cm}

\begin{large}

Correspondence to: Michail Papathomas

\end{large}

{\it The Observatory, University of St Andrews, Buchanan Gardens, St Andrews, Fife, Scotland, UK, KY16 9LZ, michail@mcs.st-and.ac.uk}

\vspace{0.5cm}

\noindent {\bf Appendix.}
\vspace{0.1cm}
{\it Proof of Theorem 1:} Assume that the subjects are grouped into $C$ clusters. As $\sum_{c=1}^{C} \gamma_{p}^{c} \times \gamma_{q}^{c} =0$, without any loss of generality, assume that, for $x_{.p}$ and $x_{.q}$, \\
 $\gamma_{p}^{c}=0$, $\gamma_{q}^{c}=1$, for $c \in \Gamma_1$, \\
 $\gamma_{p}^{c}=1$, $\gamma_{q}^{c}=0$, for $c \in \Gamma_2$, \\
 $\gamma_{p}^{c}=0$, $\gamma_{q}^{c}=0$, for $c \in \Gamma_3=\{1,\dots, C\} \cap (\Gamma_1 \cup \Gamma_2)^{\complement}$, \\
where $\Gamma_1 \cap \Gamma_2 = \emptyset$. To simplify the notation, we suppress the $x$ and $x^{'}$ from $P(x_{.p}=x, x_{.q}=x^{'})$, and write $P(x_{.p}, x_{.q})$. We also write $\phi_p^c$ instead of $\phi_{p}^{c}(x)$, and $\pi_p$ instead of $\pi_p(x)$. Finally, we write $\sum_{\Gamma_l}$, $l=1,2,3$, instead of $\sum_{c\in \Gamma_l}$. Then,
\begin{eqnarray}
P(x_{.p}, x_{.q}) &=& \sum_{c=1}^{C} \psi_c
\{(\phi_p^c)^{\gamma_p^c} (\pi_p)^{1-\gamma_p^c} \}
\{ (\phi_q^c)^{\gamma_q^c} (\pi_q)^{1-\gamma_q^c} \}  \nonumber \\
&=& \pi_p \sum_{\Gamma_1} \psi_c \phi_q^c + \pi_q \sum_{\Gamma_2} \psi_c \phi_p^c +
\pi_p \pi_q \sum_{\Gamma_3} \psi_c. \nonumber
\end{eqnarray}

Also,
\begin{eqnarray}
P(x_{.p}) P(x_{.q}) &=& \left( \sum_{\Gamma_1} \psi_c \pi_p + \sum_{\Gamma_2} \psi_c \phi_p^c + \sum_{\Gamma_3} \psi_c  \pi_p \right)
\times \left( \sum_{\Gamma_1} \psi_c \phi_q^c + \sum_{\Gamma_2} \psi_c \pi_q + \sum_{\Gamma_3} \psi_c  \pi_q  \right) \nonumber \\
&=& \pi_p \left( \sum_{\Gamma_1} \psi_c \phi_q^c \right) \left( 1- \sum_{\Gamma_2} \psi_c \right)
+ \pi_q \left( \sum_{\Gamma_2} \psi_c \phi_p^c \right) \left( 1- \sum_{\Gamma_1} \psi_c \right) 
\nonumber \\
&+& \pi_p \pi_q \left\{ \left( \sum_{\Gamma_1} \psi_c \right)
\left( \sum_{\Gamma_2} \psi_c \right) + \left( \sum_{\Gamma_3} \psi_c \right) \right\}
+ \left( \sum_{\Gamma_2} \psi_c \phi_p^c \right) \left( \sum_{\Gamma_1} \psi_c \phi_q^c \right). 
\nonumber 
\end{eqnarray}

Now,
\[
P(x_{.p}, x_{.q}) - P(x_{.p}) P(x_{.q}) = 0
\]
\[
\Leftrightarrow \pi_p \left( \sum_{\Gamma_1} \psi_c \phi_q^c \right) \left( \sum_{\Gamma_2} \psi_c \right)
+ \pi_q \left( \sum_{\Gamma_2} \psi_c \phi_p^c \right) \left( \sum_{\Gamma_1} \psi_c \right)
\]
\[
- \pi_p \pi_q \left\{ \left( \sum_{\Gamma_1} \psi_c \right) \left( \sum_{\Gamma_2} \psi_c \right) \right\}
- \left( \sum_{\Gamma_2} \psi_c \phi_p^c \right) \left( \sum_{\Gamma_1} \psi_c \phi_q^c \right) = 0
\]
\[
\Leftrightarrow
\left\{ \pi_p \left( \sum_{\Gamma_2} \psi_c \right) - \sum_{\Gamma_2} \psi_c \phi_p^c \right\}
\left\{ \sum_{\Gamma_1} \psi_c \phi_q^c - \pi_q \left( \sum_{\Gamma_1} \psi_c \right) \right\}=0
\]
This is always true since, for example, as $\pi_{p}(x)=P(x_{.p}=x)$,
\[
\pi_p=P(x_{.p}) = \sum_{c} \psi_c (\phi_p^c)^{\gamma_p^c} (\pi_p)^{1-\gamma_p^c} =
\pi_p \sum_{\Gamma_1 \cup \Gamma_3} \psi_c +   \sum_{\Gamma_2} \psi_c \phi_p^c
\]
\[
\Rightarrow \sum_{\Gamma_2} \psi_c \phi_p^c = \pi_p - \pi_p \left( 1-\sum_{\Gamma_2} \psi_c \right)
\]
\[
\Rightarrow \sum_{\Gamma_2} \psi_c \phi_p^c = \pi_p \sum_{\Gamma_2} \psi_c .
\]

\vspace{0.1cm}
{\it Proof of Theorem 2:} Without loss of generality, to simplify the notation assume that $p=1$. Then, for all $q\in \{2,\dots,\dots,P \}$, $\sum_{c=1}^{C} \gamma_{1}^{c} \times \gamma_{q}^{c} =0$.  
From Theorem 1, $x_{.1}$ is independent of $x_{.q}$, for any $2\leq q \leq P$. Such pairwise independence does not imply that $x_{.1}$ is independent of $\{x_{.2},\dots,x_{.P} \}$. To show this assume, also without loss of generality,  that $\gamma_{1}^{c}=0$, for $c \in \Gamma_1$ and $\gamma_{1}^{c}=1$, for $c \in \Gamma_2$. The $\Gamma_1$ and $\Gamma_2$ sets can be empty. Now, since, $\sum_{c=1}^{C} \gamma_{1}^{c} \times \gamma_{q}^{c} =0$, for all $q\in \{2,\dots,\dots,P \}$, $\gamma_{q}^{c}=0$ for all $c\in \Gamma_2$.  Then, 
\[
P(x_{.1},x_{.2},\dots,x_{.P}) = \sum_{c=1}^{C} \psi_c
\{(\phi_1^c)^{\gamma_1^c} (\pi_1)^{1-\gamma_1^c} \} \times 
\{ (\phi_2^c)^{\gamma_2^c} (\pi_2)^{1-\gamma_2^c} \} \times \dots \times 
\{ (\phi_P^c)^{\gamma_P^c} (\pi_P)^{1-\gamma_P^c} \}
\]
\[
= \pi_1 \times \sum_{\Gamma_1} \psi_c 
\{ (\phi_2^c)^{\gamma_2^c} (\pi_2)^{1-\gamma_2^c} \} \times \dots \times 
\{ (\phi_P^c)^{\gamma_P^c} (\pi_P)^{1-\gamma_P^c} \}
\]
\[
+ \pi_1 \times \pi_2 \times \dots \times \pi_P \times \sum_{\Gamma_2} \psi_c \phi_1^c
\]
Now,
\[
\pi_1 = \sum_{c\in \Gamma_1 \cup \Gamma_2} \psi_c (\phi_1^c)^{\gamma_1^c} (\pi_1)^{1-\gamma_1^c} =
\pi_1 \sum_{\Gamma_1} \psi_c +   \sum_{\Gamma_2} \psi_c \phi_1^c 
\]
\[
\Rightarrow \sum_{\Gamma_2} \psi_c \phi_1^c = \pi_1 - \pi_1 \left( 1-\sum_{\Gamma_2} \psi_c \right)
\Rightarrow \sum_{\Gamma_2} \psi_c \phi_1^c = \pi_1 \sum_{\Gamma_2} \psi_c,
\]

Therefore,
\[
P(x_{.1},x_{.2},\dots,x_{.P})
\]
\[
= \pi_1 \left( \sum_{\Gamma_1} \psi_c 
\{ (\phi_2^c)^{\gamma_2^c} (\pi_2)^{1-\gamma_2^c} \} \times \dots \times 
\{ (\phi_P^c)^{\gamma_P^c} (\pi_P)^{1-\gamma_P^c} \} + 
\pi_2 \times \dots \times \pi_P \times \sum_{\Gamma_2} \psi_c 
\right)
\]
\[
= P(x_{.1})\times P(x_{.2},\dots,x_{.P}),
\]

and $x_{.1}$ is independent of $\{x_{.2},\dots,x_{.P} \}$ as required. 

\newpage
\pagebreak

\begin{small}
\begin{center}
\begin{table*}[!t]
\begin{tiny}
{\bf Table 1a:} {Cluster profiles in hypothetical simple illustration, defined by the $\phi_{p}^{c}$ multinomial probabilities, for covariate $x_{.p}$ and cluster $c$.}
\label{tab:5}
\par
\begin{tabular}{ccccccc}
\hline
  & $x_{.1}$ & $x_{.2}$ & $x_{.3}$ & $x_{.4}$ & $x_{.5}$ & $x_{.6}$  \\
\hline
Cluster 1 & $(0.01,0.3,0.69)$ & $(0.01,0.3,0.69)$ & $(0.1,0.1,0.8)$ & $(0.1,0.1,0.8)$ & $(0.8,0.1,0.1)$ & $(0.8,0.1,0.1)$  \\
Cluster 2 & $(0.01,0.5,0.49)$ & $(0.01,0.5,0.49)$ & $(0.8,0.1,0.1)$ & $(0.8,0.1,0.1)$ & $(0.8,0.1,0.1)$ & $(0.8,0.1,0.1)$  \\
Cluster 3 & $(0.29,0.7,0.01)$ & $(0.29,0.7,0.01)$ & $(0.8,0.1,0.1)$ & $(0.8,0.1,0.1)$ & $(0.8,0.1,0.1)$ & $(0.8,0.1,0.1)$  \\
\hline 
\end{tabular}
\end{tiny}
\end{table*}
\end{center}
\end{small}

\begin{small}
\begin{center}
\begin{table*}[!t]
\begin{tiny}
{\bf Table 1b:} {Summary cluster profiles in hypothetical simple illustration. The `$<$' (`$>$') symbol denotes that observation $x$ of covariate $x_{.p}$ in cluster $c$ is more (less) likely compared to the average in the whole sample; otherwise, the `0' symbol is used.}
\label{tab:5}
\par
\begin{tabular}{ccccccc}
\hline
  & $x_{.1}$ & $x_{.2}$ & $x_{.3}$ & $x_{.4}$ & $x_{.5}$ & $x_{.6}$  \\
\hline
$\mbox{Median}(\rho_{p})$ & 0.8 & 0.8 & 0.9 & 0.9 & 0.001 & 0.001 \\
Cluster 1 & $<<>$ & $<<>$ & $<0>$ & $<0>$ & $000$ & $000$  \\
Cluster 2 & $<0>$ & $<0>$ & $>0<$ & $>0<$ & $000$ & $000$  \\
Cluster 3 & $>><$ & $>><$ & $>0<$ & $>0<$ & $000$ & $000$  \\
\hline 
\end{tabular}
\end{tiny}
\end{table*}
\end{center}
\end{small}

\begin{small}
\begin{center}
\begin{table*}[!h]
\begin{tiny}
{\bf Table 2:} {Simulation specifications.}
\par
\begin{tabular}{lcccccc}
\hline
 & Number & Number & Number of levels & Number of cells & Approximate number  & Number of covariates\\
 & of subjects  & of covariates & of covariates & in contingency table & of models & that form interactions \\
Simulation 1 & 10000 & 10  & 2 & 1024 & $3.5184 \times 10^{13}$ & 7 \\
Simulation 2 & 10000 & 10  & 2 & 1024 & $3.5184 \times 10^{13}$ & 6 \\
Simulation 3 & 10000 & 10  & 2 & 1024 & $3.5184 \times 10^{13}$ & 9 \\
Simulation 4 & 5000 & 20  & 3 & $3.4\times 10^{9}$ & $1.5 \times 10^{57}$ & 6 \\
Simulation 5 & 10000 & 100  & 2 & $1.27\times 10^{30}$ & $2^{4950}$ & 8 \\
\hline
\end{tabular}
\end{tiny}
\end{table*}
\end{center}
\end{small}

\begin{small}
\begin{center}
\begin{table*}[!h]
\begin{tiny}
{\bf Table 3:} {MCMC specifications for the clustering analyses, and also for the log-linear model comparison Reversible jump chains. Clustering analyses were performed using the R package PReMiuM. Reversible jump analyses were performed using Matlab code. All analyses performed on a PC equipped with an Intel(R) Core(TM)i7-2600K CPU 3.40 GHz with 8GB RAM}
\par
\begin{tabular}{lcccc}
\hline
\multicolumn{5}{c}{Clustering algorithms} \\
 & Burn-in & Iterations after burn-in & Run time in minutes (approx.) & Comment \\
Simulation 1 & 40000 & 20000  & 24 & \\
Simulation 2 & 40000 & 20000  & 24 & \\
Simulation 3 & 40000 & 20000 &  24 & \\
Simulation 4 & 100000 & 20000 & 30 & \\ 
Simulation 5 & 100000 & 20000 & 90 &      \\
Edwards and Havranek data (CHD) & 40000 & 20000 & 3 &      \\
Genetic-environmental data & 40000 & 20000 & 10 &      \\
\hline
\multicolumn{5}{c}{Reversible jump chains}  \\
 & Burn-in & Iterations & Run time in minutes  & Comment \\
Simulation 1 & 10000 & 100000  & 420 & \\
Simulation 2 & 10000 & 100000  & 420 & \\
Simulation 3 & 10000 & 100000 &  420 & \\
Simulation 4 & 2000 & 10000  & 360  & after discarding 12 covariates  \\
Simulation 5 & 50000 & $10^6$  &   240  & after discarding 92 covariates \\
Edwards and Havranek data (CHD) & 20000 & $10^6$  &   65  &  \\
Genetic-environmental data  & 20000 & $10^6$  &   65  & after discarding 18 SNPs \\
\hline
\end{tabular}
\end{tiny}
\end{table*}
\end{center}
\end{small}

\newpage
\pagebreak

\begin{small}
\begin{center}
\begin{table*}[!h]
\begin{tiny}
{\bf Table 4:} {Cluster profiles for the five simulations. In parenthesis the number of subjects typically allocated to each representative cluster. All posterior median selection probabilities for the remaining 14 covariates in Simulation 4 were less than 0.14. Posterior median selection probabilities for the remaining 92 covariates in Simulation 5 were either equal to zero or smaller than 0.01}
\label{tab:5}
\par
\begin{tabular}{lccccccccccc}
\hline 
\multicolumn{11}{c}{Simulation 1} \\
  & A & B & C & D & E & F & G & H & I & J \\
\hline
$\mbox{Median}(\rho_{p})$ & 0.36 & 0.78 & 0.32 & 0.75 & 0.06 & 0.05 & 0.00 & 0.48 & 0.57 & 0.50   \\
Cluster 1 (5465)& $><$ & $<>$ & $00$ & $<>$ & $00$ & $00$ & $00$ & $><$ & $><$  & $<>$   \\
Cluster 2 (3159)& $<>$ & $><$ & $00$ & $><$ & $00$ & $00$ & $00$ & $><$ & $<>$  & $><$   \\
Cluster 3 (1376)& $00$ & $><$ & $00$ & $00$ & $00$ & $00$ & $00$ & $<>$ & $<>$  & $<>$   \\
\hline 
\multicolumn{11}{c}{Simulation 2} \\
  & A & B & C & D & E & F & G & H & I & J \\
\hline
$\mbox{Median}(\rho_{p})$ & 0.63 & 0.38 & 0.35 & 0.53 & 0.00 & 0.50 & 0.51 & 0.16 & 0.07  & 0.09   \\
Cluster 1 (1153)& $00$ & $00$ & $00$ & $00$ & $00$ & $><$ & $><$ & $00$ & $00$  & $00$   \\
Cluster 2 (1926)& $<>$ & $><$ & $00$ & $><$ & $00$ & $><$ & $><$ & $00$ & $00$  & $00$   \\
Cluster 3 (2031)& $<>$ & $<>$ & $><$ & $><$ & $00$ & $><$ & $><$ & $00$ & $00$  & $00$   \\
Cluster 4 (2466)& $<>$ & $><$ & $<>$ & $<>$ & $00$ & $><$ & $><$ & $00$ & $00$  & $00$   \\
Cluster 5 (2424)& $><$ & $<>$ & $00$ & $<>$ & $00$ & $<>$ & $<>$ & $00$ & $00$  & $00$   \\
\hline 
\multicolumn{11}{c}{Simulation 3} \\
  & A & B & C & D & E & F & G & H & I & J \\
\hline
$\mbox{Median}(\rho_{p})$ & 0.38 & 0.50 & 0.30 & 0.54 & 0.07 & 0.34 & 0.49 & 0.41 & 0.43  & 0.66   \\
Cluster 1 (7676)& $<>$ & $><$ & $00$ & $><$ & $00$ & $><$ & $><$ & $00$ & $00$  & $00$   \\
Cluster 2 (2324)& $><$ & $<>$ & $00$ & $<>$ & $00$ & $<>$ & $<>$ & $00$ & $00$  & $00$   \\
\hline 
\multicolumn{11}{c}{Simulation 4} \\
  & A & B & C & D & E & F & & & & \\
\hline
$\mbox{Median}(\rho_{p})$ & 0.92 & 0.87 & 0.97 & 0.56 & 0.70 & 0.46   & & & &   \\
Cluster 1 (2986)& $><>$ & $><>$ & $><>$ & $><>$ & $><>$ & $><>$  & & & &   \\
Cluster 2 (306) & $000$ & $<><$ & $<><$ & $0>0$ & $<00$ & $000$  & & & &   \\
Cluster 3 (700) & $><>$ & $><0$ & $><>$ & $<><$ & $<><$ & $<><$  & & & &   \\
Cluster 4 (260) & $<><$ & $<><$ & $<><$ & $<><$ & $<><$ & $<><$  & & & &   \\ 
Cluster 5 (354) & $<><$ & $<><$ & $00>$ & $000$ & $0><$ & $000$  & & & &   \\ 
Cluster 6 (394) & $<><$ & $0>0$ & $<><$ & $000$ & $0<>$ & $0><$  & & & &   \\ 
\hline
\multicolumn{11}{c}{Simulation 5} \\ % simulation 6
  & A & B & C & D & E & F & G & H  \\
\hline
$\mbox{Median}(\rho_{p})$ & 0.96 & 0.95 & 0.97 & 0.93 & 0.97 & 0.96 & 0.97 & 0.96    \\
Cluster 1 (4036)& $><$ & $<>$ & $<>$ & $<>$ & $<>$ & $><$ & $<>$ & $<>$    \\
Cluster 2 (3813)& $><$ & $<>$ & $<>$ & $<>$ & $><$ & $<>$ & $><$ & $><$   \\
Cluster 3 (399)& $><$ & $00$ & $<>$ & $><$ & $<>$ & $><$ & $><$ & $<>$    \\
Cluster 4 (720)& $<>$ & $><$ & $><$ & $><$ & $<>$ & $><$ & $<>$ & $<>$    \\
Cluster 5 (902)& $<>$ & $><$ & $><$ & $><$ & $><$ & $<>$ & $><$ & $><$   \\
Cluster 5 (130)& $<>$ & $><$ & $><$ & $><$ & $><$ & $<>$ & $><$ & $<>$   \\
\hline
\multicolumn{11}{c}{Edwards and Havranek data (CHD)} \\ 
  & A & B & C & D & E & F & & & & \\
\hline
$\mbox{Median}(\rho_{p})$ & 0.86 & 0.92 & 0.94 & 0.26 & 0.81 & 0.10   & & & &   \\
Cluster 1 (900)& $><$ & $<>$ & $><$ & $00$ & $<>$ & $00$  & & & &   \\
Cluster 2 (941)& $<>$ & $><$ & $<>$ & $00$ & $><$ & $00$  & & & &   \\
\hline
\multicolumn{11}{c}{Genetic-environmental data (GE)} \\
  & rs8034191 (A) & rs4324798 (B) & rs1950081 (C) & age (D) & sex (E) & smoking (F) & & & & \\
  \hline
$\mbox{Median}(\rho_{p})$ & 0.01 & 0.00 & 0.10 & 0.92 & 0.82 & 0.85   & & & &   \\
Cluster 1 (2222) & $00$ & $00$ & $00$ & $><$ & $><$ & $<>$  & & & &   \\
Cluster 2 (2059) & $00$ & $00$ & $00$ & $<>$ & $<>$ & $><$  & & & &   \\
\end{tabular}
\end{tiny}
\end{table*}
\end{center}
\end{small}

\newpage
\pagebreak

\begin{small}
\begin{center}
\begin{table*}[!t]
\begin{tiny}
{\bf Table 5:} {Mixing performance of samplers. Median of iterations to best model is calculated after 30 runs of the reversible jump MCMC chain. First and third quartiles are given in parentheses. PDV denotes the unrefined model search strategy adopted in Papathomas et al (2011b). See Figure 2 for the highest posterior probability model.}
\par
\begin{tabular}{lccc}
\hline
\multicolumn{4}{c}{Simulation 1} \\
 & Acceptance rate & Iterations (median) to highest& Posterior probability \\
 &  as a percentage & posterior probability model & for highest probability model  \\
(a) Uniformly random (PDV) & 5.1 & 590 (452,821)  & 0.55 \\
(b) Cluster specific & 3.8 & 247 (164,369) & 0.55 \\
(c) Combined (30\%,10\%) & 5.3 & 540 (290,674)  & 0.53 \\
(d) Combined (20\%,20\%) & 4.9 & 403 (312,493)  & 0.55 \\
\hline
\multicolumn{4}{c}{Simulation 2} \\
 & Acceptance rate & Iterations (median) to highest& Posterior probability \\
 &  as a percentage & posterior probability model & for highest probability model \\
(a) Uniformly random (PDV) & 4.4 & 717 (475,990)  & 0.60 \\
(b) Cluster specific & 4.4 & 189 (147,238)  & 0.58 \\
(c) Combined (30\%,10\%) & 4.4 & 417 (346,354)  & 0.60 \\
(d) Combined (20\%,20\%) & 4.5 & 257 (181,314)  & 0.59 \\
\hline
\multicolumn{4}{c}{Simulation 3} \\
 & Acceptance rate & Iterations (median) to highest& Posterior probability \\
 &  as a percentage & posterior probability model & for highest probability model  \\
(a) Uniformly random (PDV) & 3.2 & 657 (545,1065)  & 0.62 \\
(b) Cluster specific & 3.1 & 445 (335,592)  & 0.60 \\
(c) Combined (30\%,10\%) & 3.3 & 538 (431,701)  & 0.60 \\
(d) Combined (20\%,20\%) & 3.2 & 560 (368,815)  & 0.61 \\
\hline
\multicolumn{4}{c}{Simulation 4 (considering only the 6 important covariates)} \\
 & Acceptance rate & Iterations (median) to highest& Posterior probability \\
 &  as a percentage & posterior probability model & for highest probability model  \\
(a) Uniformly random & 2.2 & 661 (550,746)  & 0.55 \\
(b) Cluster specific & 2.08 & 685 (534,1015)  & 0.49 \\
(c) Combined (30\%,10\%) & 2.5 & 625 (543,806)  & 0.42 \\
(d) Combined (20\%,20\%) & 2.2 & 733 (551,947)  & 0.62 \\
\hline
\multicolumn{4}{c}{Simulation 5 (considering only the 8 important covariates)} \\ %SIM 6
 & Acceptance rate & Iterations (median) to highest& Posterior probability \\
 &  as a percentage & posterior probability model & for highest probability model \\
Any of the 4 equivalent strategies & 1.1 & 5183 (3711,6590)  & 0.74 \\
\hline
\end{tabular}
\end{tiny}
\end{table*}
\end{center}
\end{small}

\begin{small}
\begin{center}
\begin{table*}[!t]
\begin{tiny}
{\bf Table 6:} {Mixing performance of samplers. Median of iterations to best model is calculated after 300 runs of the reversible jump MCMC chain. First and third quartiles are given in parentheses. PDV denotes the unrefined model search strategy adopted in Papathomas et al (2011b).}
\par
\begin{tabular}{lccc}
\hline
\multicolumn{4}{c}{Edwards and Havranek data (CHD)} \\
 & Acceptance rate & Iterations (median) to highest& Posterior probability \\
 &  as a percentage & posterior probability model & for highest probability model \\
 & &  &  `ADE+AC+BC+BE+F' \\
(a) Uniformly random (PDV) & 5.2 & 314 (215,582)  & 0.28 \\
(b) Cluster specific & 3.7 & 244 (162,378)  & 0.28 \\
(c) Combined (30\%,10\%) & 4.9 & 273 (172,470)  & 0.27 \\
(d) Combined (20\%,20\%) & 4.6 & 248 (155,392)  & 0.28 \\
\hline
\multicolumn{4}{c}{Genetic-environmental data [including important (characterized as such by clustering) representative SNPs]} \\
 & Acceptance rate & Iterations (median) to highest& Posterior probability \\
 &  as a percentage & posterior probability model & for highest probability model \\
 & &  &  `A+B+C+DEF' \\
(a) Uniformly random & 6.3 & 564 (257,1205)  & 0.53 \\
(b) Cluster specific & 8.4 & 196 (83,443)  & 0.51 \\
(c) Combined (30\%,10\%)  & 6.9 & 310 (147,670)  & 0.51 \\
(d) Combined (20\%,20\%)  & 7.5 & 235 (91,516)  & 0.52 \\
\hline
\end{tabular}
\end{tiny}
\end{table*}
\end{center}
\end{small}

\newpage
\pagebreak

\begin{figure*}[!t]
\includegraphics[height=120mm, width=180mm]{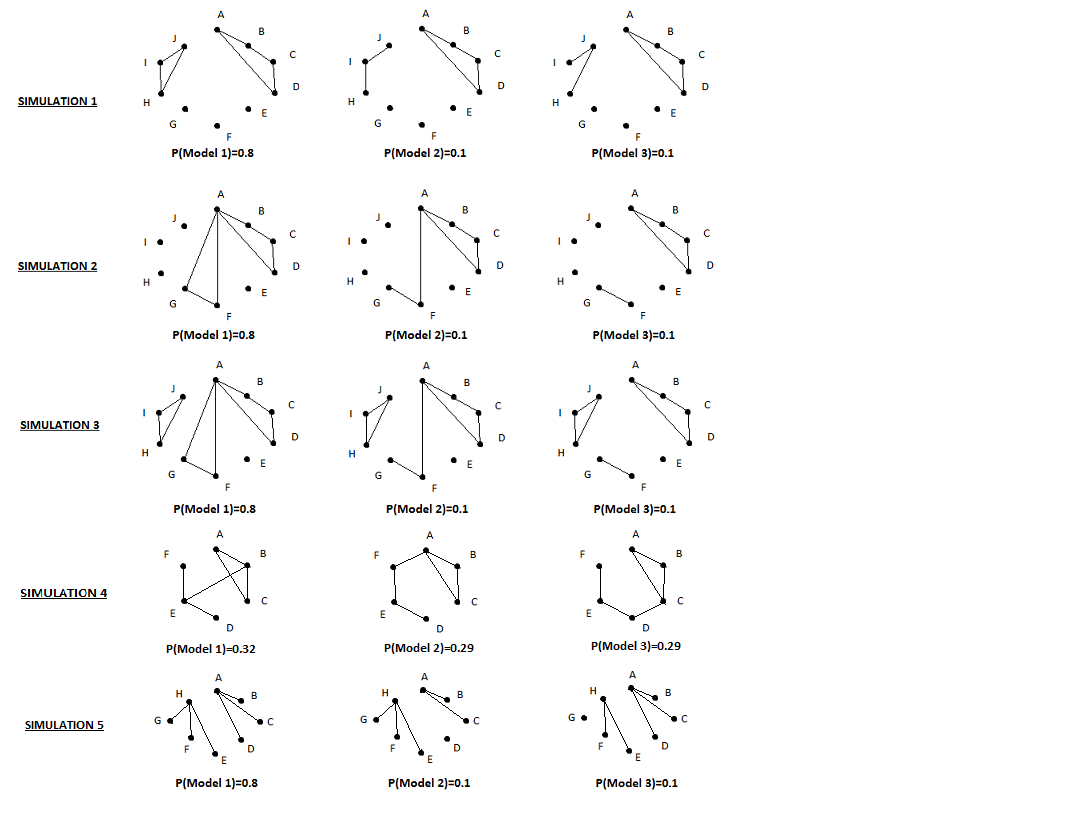}
\caption{The graphical models used for the five simulations}
\label{fig:1}       % Give a unique label
\end{figure*}

\newpage
\pagebreak

\begin{figure*}[!t]
\includegraphics[height=120mm, width=180mm]{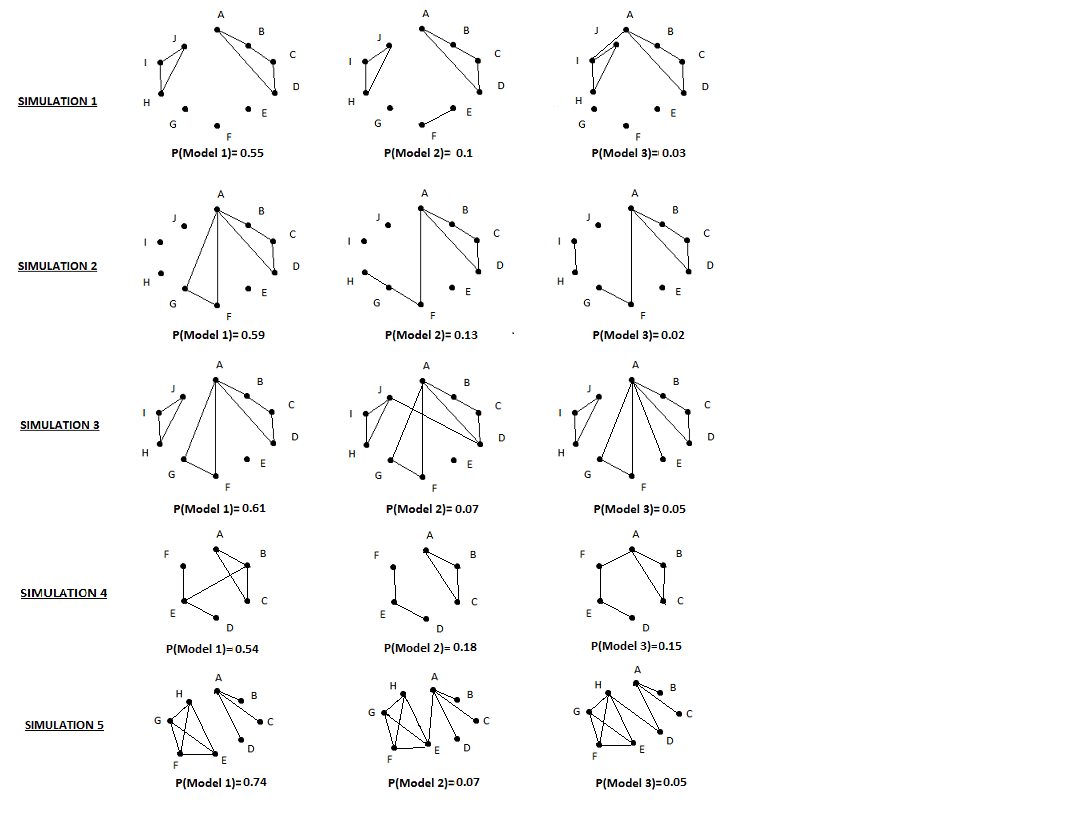}
\caption{The resulting best models from the five simulations}
\label{fig:2}       % Give a unique label
\end{figure*}

\end{document}